# Statistical details of the default priors in the Bambi library

Jacob Westfall
University of Texas at Austin

**Abstract**

This is a companion paper to Yarkoni and Westfall (2017), which describes the Python package Bambi for estimating Bayesian generalized linear mixed models using a simple interface. Here I give the statistical details underlying the default, weakly informative priors used in all models when the user does not specify the priors. Our approach is to first deduce what the variances of the slopes would be if we were instead to have defined the priors on the partial correlation scale, and then to set independent Normal priors on the slopes with variances equal to these implied variances. Our approach is similar in spirit to that of Zellner's *g*-prior (Zellner 1986), in that it involves a multivariate normal prior on the regression slopes, with a tuning parameter to control the width or informativeness of the priors irrespective of the scales of the data and predictors. The primary differences are that here the tuning parameter is directly interpretable as the standard deviation of the distribution of plausible partial correlations, and that this tuning parameter can have different values for different coefficients. The default priors for the intercepts and random effects are ultimately based on the prior slope variances.

Python library (by Yarkoni & Westfall): https://github.com/bambinos/bambi

# Introduction

Consider a regression equation:

$$Y = \beta_0 + \beta_1 X_1 + \ldots + \beta_p X_p + e.$$

Our goal is to devise a set of "weakly informative" default priors for the regression coefficients that—in the absence of any additional information supplied by the user—will still allow one to obtain reasonable parameter estimates *in general*. One possible solution would be to just use, say, standard Normal distributions for all the priors, i.e.:

$$\beta_j \sim N(0, 1).$$

This solution is obviously unsatisfactory, as it ignores the fact that the observed variables may all be on wildly different scales. So for some predictors the prior may be extremely narrow or "informative," shrinking estimates strongly toward 0, while for other predictors the prior may be extremely wide or "vague," leaving the estimates essentially unchanged.

One remedy to this differential shrinkage issue could be to set the standard deviation of the prior to a very large value, so that the prior is likely to be wide relative to almost any predictor variables. For example,

$$\beta_j \sim N(0, 10^{10}).$$

This is better, although in principle it suffers from the same problem—that is, a user could still conceivably use variables for which even this prior is implausibly narrow (although in practice that would be unlikely) or unnecessarily wide. A different but related worry is that different scalings of the variables in the same dataset—for example, due to changing the units of measurement—will lead to the priors having different levels of informativeness. This seems undesirable because scaling and shifting the variables has no meaningful consequence for traditional test statistics and standardized effect sizes (with some obvious exceptions pertaining to intercepts and models with interaction terms).

The approach we take for the default priors in Bambi is to first deduce what the variances of the slopes would be if we were instead to have defined the priors on the partial correlation scale, and then to set a Normal prior on the slope with variance equal to this implied variance. Note that technically we are *not* setting the prior directly on the partial correlation coefficient; rather, the prior is set directly on the regression coefficient, but the variance of that prior is based on a calculation of what the prior variance of the slope would be *if* we had defined the prior on the partial correlation scale. Overall, our approach is similar to that of Zellner's *g*-prior (Zellner 1986), in that it involves a multivariate normal prior on the regression slopes, with a tuning parameter to control the width or informativeness of the priors irrespective of the scale of the data. The primary difference is that in our case, the tuning parameter is directly interpretable as the standard deviation of the plausible distribution of partial correlation coefficients, and this tuning parameter can have different values for different coefficients.

To illustrate the idea, consider the case of a regression with a Normal response. One can transform the multiple regression coefficient for the predictor $X_j$ into its corresponding partial correlation (i.e., the partial correlation between the outcome and the predictor, controlling for all the other predictors) using the identity:

$$\rho_j = \beta_j \sqrt{\frac{(1-R^2_{X_j X_{-j}})\text{var}(X_j)}{(1-R^2_{Y X_{-j}})\text{var}(Y)}}, \qquad (1)$$

where $\beta_j$ is the slope for $X_j$, $R^2_{X_j X_{-j}}$ is the $R^2$ from a regression of $X_j$ on all the other predictors (ignoring $Y$), and $R^2_{Y X_{-j}}$ is the $R^2$ from the regression of $Y$ on all the predictors *other than* $X_j$.

Now suppose we were to define a prior distribution not on $\beta_j$, but rather on $\rho_j$. Let this prior have mean zero and standard deviation $\sigma_\rho$. This would imply that

$$\begin{aligned}\text{var}(\beta_j) &= \text{var}(\rho_j \sqrt{\frac{(1-R^2_{Y X_{-j}})\text{var}(Y)}{(1-R^2_{X_j X_{-j}})\text{var}(X_j)}}) \\ &= \frac{(1-R^2_{Y X_{-j}})\text{var}(Y)}{(1-R^2_{X_j X_{-j}})\text{var}(X_j)}\sigma_\rho^2\end{aligned}. \qquad (2)$$

Using this result, we could simply define a Normal prior on $\beta_j$ with mean zero and variance equal to Equation 2. In this way we can regard $\sigma_\rho$ as a tuning parameter allowing us to vary the width of the slope prior in an intuitive way.

A major limitation of this simple illustration is that it only works for models with a Normal response, since that is the only case in which Equation 1 holds. But we can extend this basic idea to work for generalized linear models (GLMs) by (1) defining a generalized partial correlation measure, (2) finding the equation of a regression coefficient in a GLM in terms of this generalized partial correlation, and (3) taking the variance of this equation.

**Generalized partial correlation**

Our generalized partial correlation coefficient is a straightforward extension of the generalized $R^2$ of Cox and Snell (1989):

$$R^2 = 1 - \exp\left(\frac{2\log(\lambda)}{n}\right),$$

where $n$ is the sample size and $\log(\lambda)$ is the difference in log-likelihoods between the estimated model and the intercept-only model. To form our generalized partial correlation coefficient for a particular predictor $X_j$, we first replace $\log(\lambda)$ with $\log(\lambda_j)$, the difference in log-likelihoods between the estimated model and the model that only omits $X_j$. Then we take the square root, with the sign of the equation determined by the sign of the corresponding regression coefficient $\beta_j$, yielding

$$\rho_j = \frac{\beta_j}{|\beta_j|}\sqrt{1 - \exp\left(\frac{2\log(\lambda_j)}{n}\right)} \qquad (3)$$

for $\beta_j \neq 0$ (and defined to be 0 when $\beta_j = 0$). This has the nice property that it is equivalent to the classical partial correlation coefficient for multiple regression models with a Normal response, but it also extends naturally to other GLMs.

**Regression coefficient in terms of the partial correlation**

The next step is to find the equation of $\beta_j$ in terms of $\rho_j$. Since the $\beta_j/|\beta_j|$ term (i.e., the sign function) simply reduces to either 1 or -1, this involves writing $\log(\lambda_j)$ in terms of $\beta_j$ and then solving for $\beta_j$. To begin, we assume that the log-likelihood function $L(\beta_j)$ for the estimated model—as a function of $\beta_j$ and with the other parameters set to their maximizing arguments given $\beta_j$—is approximately a quartic polynomial symmetrical about $\hat{\beta}_j$, the maximum likelihood value of $\beta_j$, so that it can be written in the following form:

$$L(\beta_j) \approx a(\beta_j - \hat{\beta}_j)^4 + b(\beta_j - \hat{\beta}_j)^2 + L(\hat{\beta}_j), \qquad (4)$$

where $a$ and $b$ are parameters of the quartic function, proportional to the 4th and 2nd derivatives, respectively, of the log-likelihood function, evaluated at $\hat{\beta}_j$. In practice we find $a$ and $b$ by evaluating $L(.)$ at 4 points ranging from 0 to $\hat{\beta}_j$ and then applying linear regression.

Recall that $\log(\lambda_j)$, the log-likelihood ratio, is defined as the difference in log-likelihoods between the estimated model and the model that only omits $X_j$. This gives us

$$\log(\lambda_j) = L(\hat{\beta}_j) - L(0) \approx -a\hat{\beta}_j^4 - b\hat{\beta}_j^2. \qquad (5)$$

Viewing $a$ and $b$ as fixed and $\hat{\beta}_j$ as variable, Equation 5 says: for a fixed shape of the likelihood profile, if we assume that the likelihood is maximized by a particular value of $\hat{\beta}_j$, then what would be the corresponding log-likelihood ratio?

Now we can also solve Equation 3 for

$$\log(\lambda_j) = -n\log(1 - \rho_j^2)/2 \tag{6}$$

Setting Equation 5 (treating the approximation as an equality) equal to Equation 6, we get a quartic function of $\hat{\beta}_j$:

$$a\hat{\beta}_j^4 + b\hat{\beta}_j^2 - n\log(1 - \rho_j^2)/2 = 0$$

From the shape of the log-likelihood function for GLMs we know that $b < 0$ and that generally $a > 0$. So the shape of this quartic function dictates that there are two pairs of solutions: a pair that are closer to zero, with the same absolute value but opposite sign, and a pair further from zero, again with the same absolute value but opposite sign. The pair of solutions further from zero are simply an artifact of the quartic approximation to the log-likelihood, and have no meaningful interpretation, so we discard those solutions. Between the remaining solutions, we let the sign of $\rho_j$ dictate which one to take, so that finally solving for $\hat{\beta}_j$ gives

$$\hat{\beta}_j = \frac{\rho_j}{|\rho_j|}\sqrt{\frac{b + \sqrt{b^2 + 2an\log(1 - \rho_j^2)}}{-2a}}, \tag{7}$$

which is the equation we sought. As long as $b < 0$, this will yield a real number for $\hat{\beta}_j$.

One way to check the accuracy of Equation 7—which, recall, is based on a quartic approximation to the log-likelihood—is to plug in estimated values of $\rho_j$ from many simulated datasets and check how well the resulting values of $\hat{\beta}_j$ agree with the actually estimated regression coefficients. We ran simulations involving datasets with sample sizes of 20, 100, or 400; one, two, or three predictors; zero, moderate, or high multicollinearity among the predictors; small, medium, or large associations with the response on average; and Normal, Binomial, or Poisson response variables. We found that for all parameter regimes, the $\hat{\beta}_j$ values resulting from plugging the estimated $\rho_j$ values into Equation 7 were never further than an average of 0.17% from the estimated regression coefficients.

**Variance of the regression coefficient**

Now we have the unenviable task of taking the variance of Equation 7 as a function of a random $\rho_j$. We approach the problem by approximating the variance based on a Taylor expansion about the mean of $\rho_j$.

The basic shape of the function is that it resembles a scaled version of Fisher's $z$ transformation, with the scaling dependent on $a$ and $b$: the function has the domain $(-1, 1)$ and is quite linear within the interval $[-0.7, 0.7]$ or so, but then accelerates to $\pm\infty$ as it approaches the extremes (see Figure 1). The shape of the function suggests that a Taylor expansion of odd-numbered order is appropriate, and that for small assumed values of $\sigma_\rho = \sqrt{\text{var}(\rho_j)}$, a linear approximation would work quite well, but that for larger values of $\sigma_\rho$, low-order approximations will undershoot the correct variance. The default settings in Bambi are to use a 5th-order Taylor approximation for Normal response models and a 1st-order Taylor approximation for non-Normal response models.

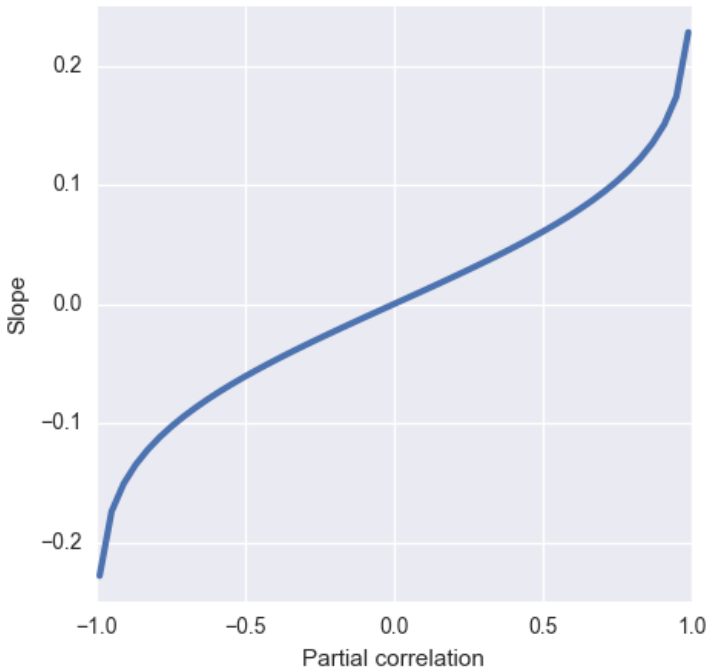

**Figure 1.** Equation 7 plotted for a range of partial correlations, for values *a* and *b* derived from a simulated example dataset.

For a function $g(X)$ of a random variable $X$, the variance of the $k$th order Taylor approximation of $g(X)$ about the point $\mu = \mathrm{E}[X]$ can be shown to be

$$\text{var}[g(X)] \approx \sum_{i=1}^{k}\sum_{j=1}^{k} \frac{1}{i!j!} g^i(\mu) g^j(\mu)(\mu^{i+j} - \mu^i \mu^j)$$
,

where $g^m(\cdot)$ is the $m$th derivative of $g(\cdot)$, and $\mu^m$ is the $m$th central moment of $X$. We do not give the derivatives of Equation 7 here, which are quite long and complicated, but the expressions can be obtained using the Python library SymPy (Meurer et al. 2017). We note that some of these derivatives do not exist at the point $\rho_j = 0$ due to the presence of the sign function in Equation 7. In these cases, for practical purposes we can obtain good results (as verified in simulations) by simply evaluating the derivatives at a point very close to 0, such as $\rho_j = .001$.

For the central moments, we have to make specific distributional assumptions about $\rho_j$ beyond just its variance. We use a Beta distribution, scaled to have support in $(-1, 1)$, with shape parameters $p$ and $q$. Then the $m$th central moment for this distribution is given by

$$_2F_1\left(p, -m; p+q; \frac{p+q}{p}\right)\left(\frac{2p}{p+q}\right)^m,$$

where $_2F_1$ is a hypergeometric function (Weisstein 2017).

Applying this method results in an implied prior variance of $\beta_j$ if we had defined the prior on the $\rho_j$ scale. We emphasize again that we do not actually place the priors directly on the $\rho_j$. Rather, we place Normal priors directly on the regression coefficients $\beta_j$, and we simply set the variances of these Normal priors to the values resulting from the process above. The result is that the slopes have the Normal priors that are familiar to most users, but now we can tune the width or informativeness of these priors in an intuitive way by setting different values of $\sigma_\rho$, corresponding to different standard deviations of the distribution of plausible partial correlations.

If no value of $\sigma_\rho$ is specified by the user, then the default value is set to $\sigma_\rho = \sqrt{1/3} \approx .577$, which is the standard deviation of a flat prior in the interval [-1,1]. Simulations show that at this default value, the 5th-order Taylor approximation results in only a slight underestimation of the correct standard deviation, corresponding to an actual standard deviation of about 0.53 or so on the $\rho_j$ scale (see Figure 2). We allow users to specify their priors in terms of the exact value of $\sigma_\rho$, or in terms of four labels: "narrow" meaning $\sigma_\rho = .2$, "medium" meaning $\sigma_\rho = .4$, "wide" meaning $\sigma_\rho = \sqrt{1/3}$ (i.e., the default), or "superwide" meaning $\sigma_\rho = .8$. (Note that the maximum possible standard deviation of a distribution of partial correlations is 1, which would be a distribution with half of the values at $\rho_j = -1$ and the other half at $\rho_j = 1$.) Viewed from this partial correlation perspective, it seems hard to theoretically justify anything wider than our "wide" default, since this would correspond to something that is wider than a flat prior on the partial correlation scale. With that said, there should be no practical, technical problem in using such a wider prior, since our Taylor approximation method means that $\text{var}(\beta_j)$ does not accelerate to $\pm\infty$ as $\sigma_\rho$ approaches -1 or 1, as the real function would.

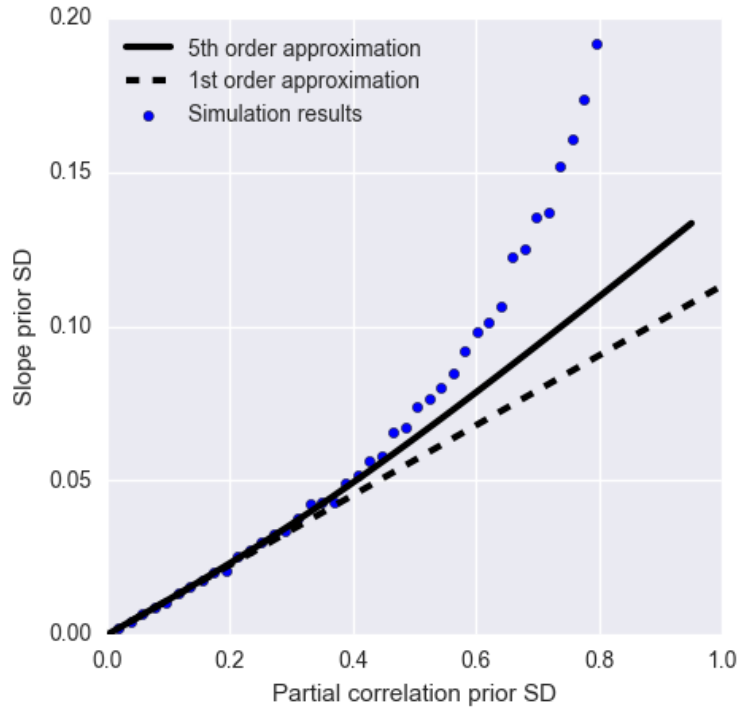

**Figure 2.** Evaluating the accuracy of the Taylor-approximate standard deviation (SD) of the slope priors by comparison to simulated datasets.

### Intercept, cell means, and residual standard deviation

The default prior for the intercept $\beta_0$ must follow a different scheme. We first note that in ordinary least squares (OLS) regression $\beta_0 = \bar{Y} - \beta_1 \bar{X}_1 - \beta_2 \bar{X}_2 - \ldots$. So we can set the mean of the prior on $\beta_0$ to

$$\mathrm{E}[\beta_0] = \bar{Y} - \mathrm{E}[\beta_1]\bar{X}_1 - \mathrm{E}[\beta_2]\bar{X}_2 - \ldots.$$

In practice, the priors on the slopes will typically be set to have zero mean, so the mean of the prior on $\beta_0$ will typically reduce to $\bar{Y}$.

Now for the variance of $\beta_0$ we have (assuming independence of the slope priors):

$$\mathrm{var}(\beta_0) = \mathrm{var}(Y)/n + \bar{X}_1^2 \mathrm{var}(\beta_1) + \bar{X}_2^2 \mathrm{var}(\beta_2) + \ldots. \tag{8}$$

In other words, once we have defined the priors on the slopes, we can combine this with the means of the predictors to find the implied variance of $\beta_0$. Our default prior for intercepts is a Normal distribution with mean and variance defined as above, except that the $\mathrm{var}(Y)/n$ term in Equation 8 is replaced by $\mathrm{var}(Y)$, so that the intercept prior will not be too narrow when the predictors are centered and the sample size is large. In non-Normal response models, we compute $\mathrm{E}[Y]$ and $\mathrm{var}(Y)$ on the link scale by estimating a GLM with only an intercept $\beta_0$ and then setting $\mathrm{E}[Y] = \hat{\beta}_0$ and $\mathrm{var}(Y) = n\mathrm{var}(\hat{\beta}_0)$. In cell-means models (i.e., models with no constant term, but instead with $k$ dummies for $k$ groups so that the dummy coefficients estimate the cell means), the priors for the cell-mean coefficients are handled the same way as intercepts.

For the residual standard deviation $\sigma_e$, we know that necessarily $0 \leq \sigma_e \leq \sqrt{\mathrm{var}(Y)}$. Ideally we would have a prior with support bounded in $[0, \sqrt{\mathrm{var}(Y)}]$, negatively skewed with minimum value at 0 (which implies a coefficient of determination $R^2 = 1$) and maximum value at $\sqrt{\mathrm{var}(Y)}$ (which implies $R^2 = 0$). Because there is not currently such a ready-made distribution in the backend packages underlying Bambi—and because it will tend to make little difference in practice—we simply use $\sigma_e \sim \mathrm{Uniform}(0, \sqrt{\mathrm{var}(Y)})$.

### Random effects

As is customary with mixed models, random effects are assumed to be Normally distributed. The default prior variances of those Normal distributions are based on the idea that, generally speaking, the greater is the prior variance of the corresponding fixed effect coefficient, the greater should be the prior variance of the random effect variance. We implement this idea by using Half Normal distributions for the random effect standard deviations, each with parameter $\sigma$ set equal to the prior standard deviation of the corresponding fixed effect. If the fixed part of the model does not include the corresponding fixed effect, then we consider an augmented model in which the fixed part of the model *does* include the corresponding fixed effect, and we compute what would be the mean and standard deviation of the prior for this fixed effect using the methods described previously, and then set $\sigma$ equal to this implied prior standard deviation.

### Limitations and future extensions

Our default prior system is based on independent Normal priors for all slopes, so that their joint prior distribution is multivariate Normal with a diagonal covariance matrix. It is possible that allowing this multivariate Normal to have non-zero covariances that are some function of the correlations among the predictors would make sense.

Our interpretation of the tuning parameter $\sigma_\rho$ as the standard deviation of the distribution of plausible partial correlations is technically only valid for models without random effects (i.e., GLMs). For models with random effects, varying the tuning parameter within the usual range should still result in sensible and useful prior scales, but these scales cannot technically be interpreted in terms of partial correlations. It is certainly possible that our system could be extended to provide the same intuitive, correlation-based interpretation in the presence of random effects--since the generalized $R^2$ on which our partial correlation is based simply depends on the likelihoods of the models being compared, which are perfectly well-defined for mixed models--but for now this possibility remains to be explored.